\documentclass[twocolumn,prl,showpacs,unsortedaddress,superscriptaddress]{revtex4}

\usepackage{times,mathptmx}

\usepackage{epsfig,amsmath}

\begin{document}

\title{Exciton resonances quench the photoluminescence of zigzag carbon nanotubes}

\author{Stephanie Reich}
\affiliation{Department of Engineering, University of Cambridge, Trumpington Street, Cambridge CB2 1PZ, UK}

\author{Christian Thomsen}
\affiliation{Institut f\"ur Festk\"orperphysik, Technische Universit\"at Berlin, Hardenbergstr. 36, 10623 Berlin, Germany}

\author{John Robertson}
\affiliation{Department of Engineering, University of Cambridge, Trumpington Street,
Cambridge CB2 1PZ, UK}


\begin{abstract}
We show that the photoluminescence intensity of single-walled carbon nanotubes is much stronger in tubes with large chiral angles - armchair tubes - because exciton resonances make the luminescence of zigzag tubes intrinsically weak. This exciton-exciton resonance depends on the electronic structure of the tubes and is  found  more often in nanotubes of the $+1$ family. Armchair tubes do not necessarily grow preferentially with present growth techniques; they just have stronger luminescence. Our analysis allows to normalize photoluminescence intensities and find the abundance of nanotube chiralities in macroscopic samples.
\end{abstract}

\maketitle

A major challenge in research on single walled carbon nanotubes (SWNT) is to control and measure the nanotube chiral indices on bulk, macroscopic samples. The chiral index $(n,m)$ fixes the nanotube's diameter and chiral angle. These two parameters determine all properties of a nanotube, in particular its electronic structure.\cite{reichbuch} Photoluminescence (PL) from single-walled carbon nanotubes in solution decreases strongly in intensity for nanotubes with small chiral angles (zigzag tubes)\cite{bachilo02,miyauchi04,lebedkin03}. This has been interpreted as reflecting the abundance of $(n,m)$ nanotubes\cite{bachilo02}. If correct, the interpretation has far reaching consequences, because it implies that \emph{all} present growth techniques strongly favor armchair over zigzag tubes. Is, however, the luminescence cross section really independent of the chiral angle? Can we expect constant maximum luminescence intensities when comparing two nanotubes of different chirality?

In this paper we show that luminescence strongly favors a subset of nanotubes---tubes with a small ratio between their second and first transition energy. This, in particular, implies a stronger luminescence intensity for large chiral angles and nanotubes from the $-1$ family as observed experimentally.\cite{bachilo02,miyauchi04} The chirality dependence arises from an exciton-exciton resonance. It also shifts the experimental optical transition energies to the red when compared to the bare exciton energies. From the experimental data we calculate the exciton energies and maximum luminescence intensities of more than 40 nanotube types. 

\begin{figure}[b]
\centerline{
\epsfig{file=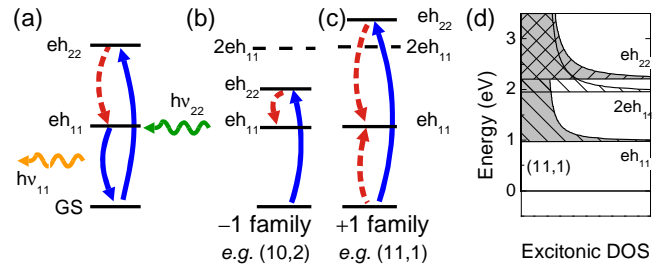,width=8.5cm,clip=}}
\caption[]{(Color online) (a) Standard picture of eh$_{22}$ excitation from the ground state (GS) by photon absorption (full arrow), exciton relaxation to eh$_{11}$ (broken), and emission (full) of $h\nu_{11}$. Carbon nanotubes fall into those (b) with only one decay channel and those (c) where eh$_{22}$ can decay into \emph{two} eh$_{11}$; (d) excitonic density of states (DOS).}
\label{schematic}
\end{figure}

We consider first luminescence in two specific nanotubes (11,1) and (10,2) with similar diameter but very different exciton behavior. Our argument is based on a key observation. The optical transition energies are often represented by the `Kataura plot' in which the energies of each subband vary roughly inversely with nanotube diameter. The transition energies deviate systematically above and below this trend, according to families of chiral indices.\cite{reichbuch,reich00c} We show that this deviation leads to an extra exciton decay channel in tubes with small band gaps.

Figure~\ref{schematic}(a) shows the PL process in SWNTs. A photon $h\nu_{22}$ creates an exciton eh$_{22}$ in the second subband of the tube, where the index 2 refers to subband 2. The exciton relaxes to the lower subband state eh$_{11}$ and recombines to the ground state, emitting the photon $h\nu_{11}$. Strong PL occurs if $h\nu_{22}$ corresponds to a singularity in the excitonic density of states, see Fig.~\ref{schematic}(d).\cite{pl}

Figure~\ref{schematic}(a) changes fundamentally if we allow the presence of two excitons in subband 1, which we denote by 2eh$_{11}$.\cite{kane03} In a (10,2) nanotube, the eh$_{22}$ energy $E_{22}$ is unusually low so the 2-exciton state lies above the eh$_{22}$ state. The standard picture is retained, Fig.~\ref{schematic}(b), and the exciton just decays into subband 1. 
On the other hand, in the (11,1) nanotube, eh$_{22}$ is high in energy and 2eh$_{11}$ lies below it. The eh$_{22}$ exciton decays into the eh$_{11}$ state with energy $E_{11}$ and thereby liberates enough energy to create a second exciton eh$_{11}$, Fig.~\ref{schematic}(c). There are two crucial points about the eh$_{22}\rightarrow$ 2eh$_{11}$ decay channel: Whether it is allowed energetically or not depends on the nanotube structure. Two seemingly similar nanotubes can show very different exciton dynamics. Second, the 2eh$_{22}$ state has a singular energy dependence [Fig.~\ref{schematic}(d)]; therefore the higher-order process strongly affects the nanotube optical properties.

\begin{figure}
\epsfig{file=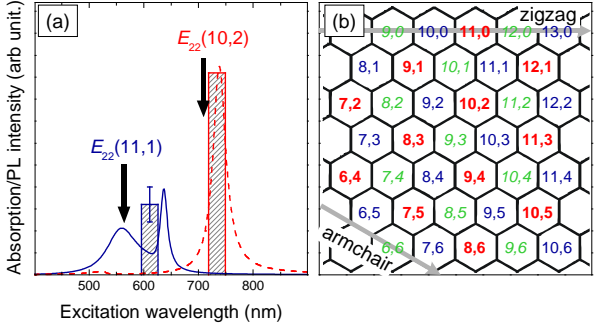,width=8.5cm,clip=}
\caption[]{(Color online) (a) Columns: excitation wavelength with the strongest PL and the intensity of tubes in bulk samples.\cite{bachilo02,miyauchi04} Lines: calculated profiles (see text). (b) Green, italic: $p=0$, red, bold: $+1$, and blue: $-1$.}
\label{absorption}
\end{figure}

In bulk SWNT samples the PL of (11,1) nanotubes is found to be three times weaker than from (10,2) nanotubes, see Fig.~\ref{absorption}(a). This factor was observed in nanotubes grown by different methods\cite{bachilo02,miyauchi04}, so it is unlikely to be due to a chirality or diameter dependence in the growth process. The (11,1) and (10,2) nanotubes have very similar diameters (9.0 and 8.7 {\AA}) and similar, small chiral angles (4.3$^\circ$ and 8.9$^\circ$). They are neighbors on the standard $(n,m)$ plot of a graphene sheet in Fig.~\ref{absorption}(b). However, one characteristic clearly distinguishes (10,2) and (11,1) tubes - they belong to different index families. 

\begin{figure}
\centerline{
\epsfig{file=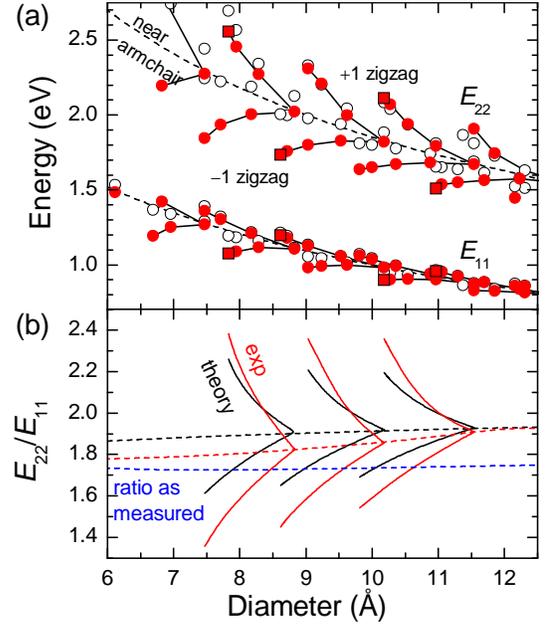,width=8.5cm,clip=}}
\caption[]{(Color online) (a) Closed symbols show the bare exciton energies; open symbols were calculated. Full lines connect tubes in a branch ($2n+m=\,$const.), dashed lines tubes with largest chiral angles (close to armchair); squares are zigzag tubes. (b) Ratio of $E_{22}$ and $E_{11}$ for three branches. The dashed blue line is the as measured ratio $h\nu_{22}/h\nu_{11}$.}
\label{kataura}
\end{figure}

SWNTs are characterized by a `family index' $p = (n-m)\,\mathrm{mod}\,3$.\cite{reich00c}
Tubes with $p = 0$ are metallic.\cite{reichbuch} Tubes with $p = +1$ or $-1$ are semiconductors. For $+1$ tubes like (11,1), the eh$_{22}$ transition energies $E_{22}$ lie above the averaged Kataura trend, for $-1$ tubes they lie below this trend, Fig.~\ref{kataura}(a). This arises from the trigonal warping of the graphene band structure.\cite{reich00c,saito00} This difference carries over into the exciton energies. Tubes with $p = +1$, like (11,1), have $E_{22} > 2E_{11}$ as in Fig.~\ref{schematic}(c), while those with $p = -1$, like (10,2), have $E_{22} < 2E_{11}$ as in Fig.~\ref{schematic}(b).  Now, nanotubes in the $+1$ family have the extra eh$_{22}$ $\rightarrow$ 2eh$_{11}$ relaxation channel, which is forbidden in $-1$ tubes. We argue that this weakens the maximum luminescence intensity in $+1$ tubes  by broadening their absorption line-width. 

To test this idea, we calculated the absorption spectrum for the eh$_{22}$ exciton by a Green function method.\cite{kane03} 
\begin{equation}
I(E)\propto -\mathrm{Im}\Bigl[\frac{1}{
E-E_{22}+\frac{A\tilde\alpha^2 E_{11}}{\sqrt{E_{11}^2-E^2}}+\frac{B^2\tilde\alpha^4}{\sqrt{4E_{11}^2-E^2}}}\Bigr]
\label{green}
\end{equation}
with a dielectric screening $\tilde\alpha=0.15$\cite{kane03,kane04}, an eh$_{22}\rightarrow\,$eh$_{11}$ coupling $A\tilde\alpha^2E_{11}\approx0.01\,$eV$^2$, and an eh$_{22}\rightarrow\,2$eh$_{11}$ coupling $B^2\tilde\alpha^4=0.1\,$eV$^2$. The parameters were adjusted to fit the electron-hole decay times in single-walled carbon nanotubes and graphite.\cite{lauret03,moos01} $E_{11}$ ($E_{22}$) is the energy of the eh$_{11}$ (eh$_{22}$) exciton.

Figure~\ref{absorption}(a) shows the calculated absorption as a function of excitation wavelength. The (10,2) absorption shows a single, slightly red-shifted Lorentzian (see arrow at $E_{22}$). However, in the (11,1) nanotube, the decay to 2eh$_{11}$ changes the absorption; the strongly red-shifted 640\,nm peak has a weak side-band at higher energies. Identifying the two narrow peaks in the calculated spectrum of the (10,2) and (11,1) tube with their measured PL excitation energy and intensity, we get excellent agreement between theory and experiment, Fig.~\ref{absorption}(a). We find that the maximum PL intensity of a (11,1) tube is intrinsically weaker than that of a (10,2) nanotube.\cite{footnote_intensity}

We now generalize our findings to arbitrary carbon nanotubes. First we calculate the bare eh$_{22}$ energies rather than the renormalized energies $h\nu_{22}$ measured by optical spectroscopy.
For $E_{11}$ we use the measured transition energies, since this exciton is not renormalized.\cite{bachilo02,lebedkin03} $E_{22}$ we find in a self-consistent routine by requiring the maximum absorption probability in Eq.~\eqref{green} to occur at $h\nu_{22}$. $h\nu_{22}$ is, in general, smaller than $E_{22}$ because of the renormalization, see Fig.~\ref{schematic} and the supplementary material.\cite{online} Additionally, we calculated $E_{22}$ within the third-order tight-binding approximation;\cite{reich02b} we used the parametrization by Kane and Mele to account for electron-electron and electron-hole interaction\cite{kane04}. 

Figure~\ref{kataura}(a) shows the bare exciton energies. Above the isotropic line (black dashed), the agreement between theory and experiment is excellent; below there are deviations of 10 - 20\,\% that arise from curvature. The agreement between theory and experiment systematically improves for large diameters and chiral angles, see Fig.~\ref{kataura}(b). For close-to-armchair tubes (dashed lines) experiment matches theory above 11\,{\AA}. In contrast, the as-measured $h\nu_{22}/h\nu_{11}$ ratio is constant and 10\,\% smaller than the exciton ratio, which is called the `ratio problem'.\cite{bachilo02,kane03}

\begin{figure}
\centerline{
\epsfig{file=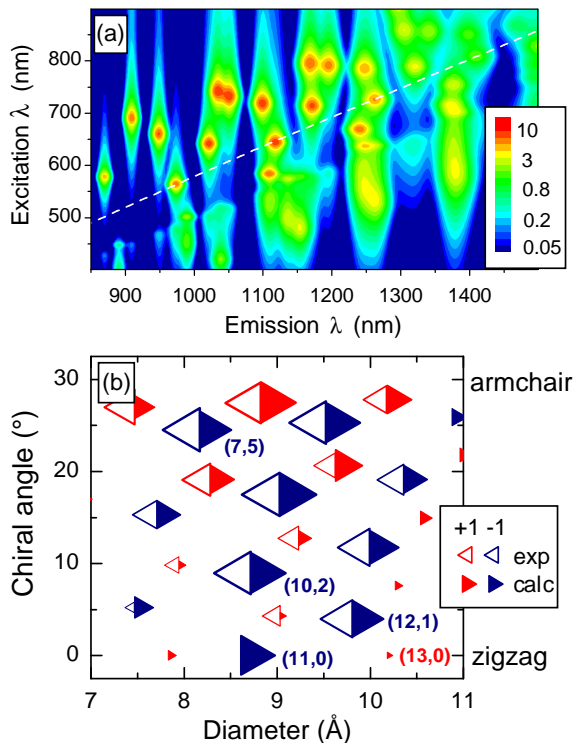,width=8.5cm,clip=}}
\caption[]{(Color online) (a) Calculated photoluminescence as a function of excitation and emission wavelength. The dashed white line shows the near-armchair direction. (b) Measured  (Fig.~3(d) in Ref.~\cite{miyauchi04}) and calcuated PL intensity for HiPCo samples as a function of nanotube diameter and chiral angle.}
\label{lumi}
\end{figure}

The PL intensity depends on the product of the absorption, relaxation, and emission probability. For a specific tube the PL intensity as a function of excitation energy simply follows Eq.~\eqref{green}. For different tubes we assume constant thermalization rates. The optical matrix elements depend on diameter as $1/d$; their dependences on chiral angle and family cancel, because of opposite trends for the second (absorption step) and first (emission) subband. The luminescence intensity hence follows the absorption probability [Eq.~(\ref{green})] weighted by $1/d^2$ and the abundance of nanotube chiralities.
Here and in the remainder of the paper we consider a simulated sample of 78 nanotubes types with mean diameter of 9.5{\AA}, a Gaussian diameter distribution of width of 2{\AA}, typical of HiPCo samples, and \emph{no} preferred chirality.

The calculated luminescence map in Fig.~\ref{lumi}(a) agrees well with the experimental false color plot of Bachilo \emph{et al.}\cite{bachilo02}. For comparison we provide a PL map for the same ensemble where we neglected the eh$_{22}\rightarrow 2$eh$_{11}$ decay [$B=0$ in Eq.~\eqref{green}] as a supplement.\cite{online} In Fig.~\ref{lumi}(a) the intensity is strong for large chiral angles in both semiconducting families. These are the PL peaks close to the armchair direction, see dashed line in Fig.~\ref{lumi}(a). Above this line is the $-1$ family of semiconducting tubes. There are 15 peaks clearly visible in this region. Below the dashed line the plot looks very different in both theory and experiment from above the line. Although there are 14 tubes in this region, we predict only four well defined peaks (compare supplement\cite{online}). They are $+1$ tubes with large chiral angles. For smaller chiral angles, trigonal warping lowers the eh$_{11}$ energy, but raises eh$_{22}$. The eh$_{22}$ $\rightarrow$ 2eh$_{11}$ resonance sets in; it blurs and broadens the absorption spectra, see Fig.~\ref{schematic}(a). In the luminescence map this creates the vertical streaks.

Figure~\ref{lumi}(b) compares the measured and calculated luminescence intensities; we see excellent agreement. The intensity of $-1$ tubes (blue) is almost independent of chiral angle. In contrast, $+1$ tubes (red) show strong luminescence for large angle, but are very weak towards the zigzag direction. In addition, luminescence hardly sees some semiconducting nanotubes at all. For example, the luminescence of a (13,0) tube (small chiral angle) is 5 times weaker than for a (7,5) tube. Thus, luminescence is strongly biased towards large chiral angles and nanotubes with $(n-m)\,\mathrm{mod}\,3=-1$.

For one nanotube, the (11,0) tube, we predicted a strong luminescence intensity although it is absent in the measured spectra [Fig.~\ref{lumi}(b)]. This could imply a small number of (11,0) nanotubes in the sample. Another explanation, however, are optically inactive excitons below eh$_{11}$.\cite{perebeinos04}. The (11,0) tube is a singular case in the experimental data as well. The two other $-1$ nanotubes with very small chiral angles [(10,2) and (12,1)] have strong intensities. Two other points are noteworthy: first, the calculations do not predicted a constant background above 1000\,nm emission wavelength. Further studies are desirable to clarify the experimental background. Second, in our calculation, there are no features below 400 and above 850\,nm excitation wavelength, because we considered only the eh$_{22}$ and the eh$_{11}$ state and their interactions.

How can we further verify the model experimentally and what are the practical implications for finding nanotube abundances from optical spectroscopy? The most rigorous test is to compare PL intensities with a chirality distribution obtained from a non-optical technique such as electron diffraction. This would establish an experimental normalization for the luminescence intensities in addition to the theoretical factors given by us.\cite{online} Time-resolved spectroscopy could be used to observe the two distinct eh$_{22}\rightarrow$ eh$_{11}$ and eh$_{22}\rightarrow 2$eh$_{11}$ decay channels. The challenge is the weak luminescence signal in tubes with the latter decay process. Another prediction from our model is a strong difference between the Raman cross section in resonance with eh$_{22}$ and eh$_{11}$.

\begin{figure}
\centerline{
\epsfig{file=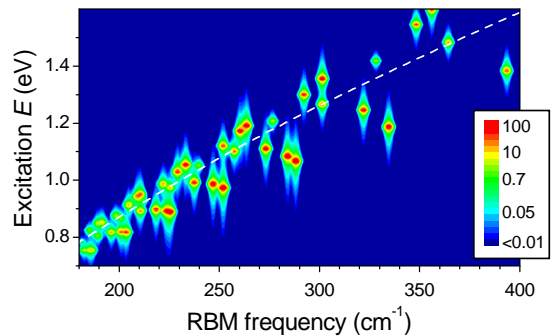,width=8.5cm,clip=}}
\caption[]{(Color online) Calculated intensity of the RBM in resonance with eh$_{11}$ for HiPCo nanotubes. The white line shows the near-armchair direction.}
\label{raman}
\end{figure}

The Raman cross section is proportional to the square of the absorption strength. Raman scattering in resonance with eh$_{22}$ therefore shows a similar depedence on tube type as photoluminescence. It is not exactly the same, because of the squared absorption probability and electron-phonon coupling, see Ref.~\cite{telg04}. For Raman scattering in resonance with eh$_{11}$, however, we predict a remarkably straightforward way to extract the chirality abundance, because there are no excitonic states below the first subband exciton.

We calculated the Raman spectra of the radial breathing mode (RBM) in resonance with eh$_{11}$ from non-orthogonal tight-binding.\cite{popov04} Although this calculation neglects electron-hole interaction, the dependence of the optical and electron-phonon matrix elements on diameter and chiral angle should be well described by a one-electron model. The Raman intensity map is shown in Fig.~\ref{raman}. In contrast to luminescence [Fig.~\ref{lumi}(a)], Raman scattering shows well resolved peaks for both semiconducting families and all chiral angles. 

The maximum Raman intensity in Fig.~\ref{raman} varies for different RBM peaks. This comes from the Gaussian diameter distribution and the dependence of the Raman matrix elements on chirality.\cite{popov04} The latter is described excellently by simple analytic functions of the nanotube diameter $d$, chiral angle $\Theta$, and family $p$. The square root of the Raman intensity follows to very good approximation (less than 10\,\% deviation)
\begin{multline}
\chi = \Bigl(1+p\,4.82\cdot10^{-2} + \frac{4.59\cdot10^{-2} d}{\mathrm{nm}}\Bigr)\Bigl[
1+(2.66\cdot10^{-3}\\+p\,6.18\cdot10^{-3})\Theta + 1.06\cdot10^{-3}\Theta^2\Bigr]\chi_0,
\label{chi}
\end{multline}
where $\Theta$ is given in degrees and $\chi_0$ is constant. Infrared Raman scattering could thus verify our model of exciton decay. Additionally, the eh$_{11}$-resonance intensities could be used to normalize PL intensities and their dependence on family and chiral angle.

In conclusion, luminescence has a systematically higher cross-section for SWNTs with large chiral angles and tubes from the $-1$ family. This arises from an additional decay channel when the exciton of the second subband has more than twice the energy of the first subband exciton. The resulting exciton-exciton resonance reduces the maximum absorption strength and also shifts the optical transition energies. As an important consequence uncorrected photoluminescence overestimates the abundance of armchair-like tubes. We suggested several experiments to verify our ideas, among them infrared Raman spectroscopy in resonance with the first subband exciton.

We thank A.C. Ferrari for a critical reading of the manuscript and useful discussions. SR was supported by the Oppenheimer Fund and Newnham College. This work was supported by DFG grant No. Th662/8-2.

%


\begin{figure*}
\epsfig{file=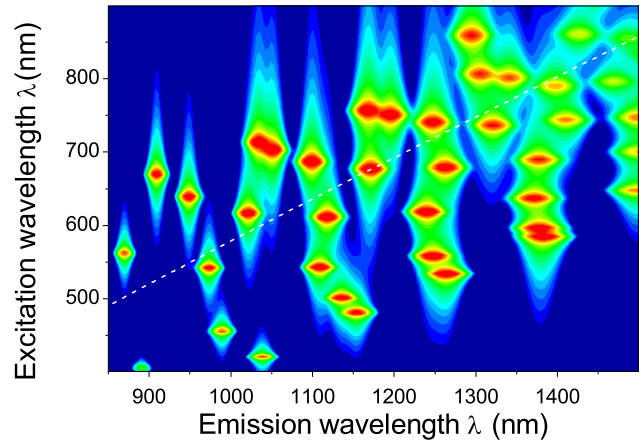,width=8.5cm,clip=}
\caption[]{Photoluminescence map calculated \emph{without} exciton-exciton resonance [$B=0$ in Eq.(1) of the paper]. The simulated sample is the same as described in the paper. This figure is to be compared with the calculation including exciton-exciton resonances [Fig.~4(a) of the main paper] and experiment [Bachilo~\emph{et al.}, Science \textbf{298}, 2361 (2002)]. Note, in particular, the many peaks below the isotropic limit (dashed line), which are no longer visible when the eh$_{22}\rightarrow 2$eh$_{22}$ decay is taken into account.}
\end{figure*}

\clearpage

\begin{table*}[h]
\caption[]{Topological and optical parameters for single-walled carbon nanotubes. $E_{11}$ and $E_{22}$ are the bare exciton energies as opposed to the measured renormalized transition energies $h\nu_{22}$. The $h\nu_{22}$ were taken from Ref.~2, 7, and 9 of the paper. PL int. (Raman int.) are the
photoluminescence [eh$_{11}$ resonant Raman, see Ref. 21 of the paper] intensities; they are normalized to the intensity of the (5,4) nanotube.
}
\begin{ruledtabular}
\begin{tabular}{lllrllllll}
\small
$(n,m)$&$d$ (nm)&$\Theta$ ($^\circ$)&$\nu$&$E_{11}$ (eV)&$E_{22}$ (eV)&$h\nu_{22}$ (eV)&PL int. &Raman int.\\&&&&&&&(arb. units)& (arb. units)\\\hline
(5,4)&	6.12&	26.3&	1&	1.48&	2.64&2.57	&100&	100\\
(6,4)&	6.83&	23.4&	-1&	1.42&	2.19&2.15	&78&	5.4\\
(6,5)&	7.47&	27&	1&	1.27&	2.27&2.19	&66&	54\\
(7,3)&	6.96&	17&	1&	1.25&	2.70&2.46	&28&	290\\
(7,5)&	8.18&	24.5&	-1&	1.21&	2.00&1.92	&55&	3.7\\
(7,6)&	8.83&	27.5&	1&	1.11&	2.02&1.91	&46&	32\\
(8,1)&	6.69&	5.8&	1&	1.19&	2.92&2.63	&26&	490\\
(8,3)&	7.72&	15.3&	-1&	1.3&	1.93&1.86	&60&	110\\
(8,4)&	8.29&	19.1&	1&	1.12&	2.27&2.11	&41&	180\\
(8,6)&	9.53&	25.3&	-1&	1.06&	1.83&1.73	&40&	2.4\\
(8,7)&	10.18&27.8&	1&	0.98&	1.82&1.70	&33&	19\\
(9,1)&	7.47&	5.2&	-1&	1.36&	1.85&1.79	&61&	290\\
(9,2)&	7.95&	9.8&	1&	1.09&	2.46&2.25	&14&	370\\
(9,4)&	9.03&	17.5&	-1&	1.13&	1.80&1.13	&44&	66\\
(9,5)&	9.63&	20.6&	1&	1.00&	2.00&1.85	&32&	110\\
(9,7)&	10.88&25.9&	-1&	0.94&	1.68&1.56	&30&	1.8\\
(9,8)&	11.54&28.1&	1&	0.88&	1.67&1.53	&25&	11\\
(10,0)&	7.83&	0&	1&	1.07&	2.56&2.31	&17&	440\\
(10,2)&	8.72&	8.9&	-1&	1.18&	1.76&1.68	&47&	200\\
(10,3)&	9.24&	12.7&	1&	0.99&	2.21&1.96	&15&	250\\
(10,5)&	10.36&19.1&	-1&	0.99&	1.67&1.57	&33&	43\\
(10,6)&	10.97&21.8&	1&	0.90&	1.79&1.64	&25&	67\\
(10,8)&	12.24&26.3&	-1&	0.84&	1.57&1.43	&22&	1.4\\
(10,9)&	12.9&	28.3&	1&	0.79&	1.55&1.40	&19&	7.1\\
(11,0)&	8.62&	0&	-1&	1.20&	1.73&1.67	&47&	270\\
(11,1)&	9.03&	4.3&	1&	0.98&	2.31&2.03	&11&	360\\
(11,3)&	10.00&11.7&	-1&	1.04&	1.65&1.56	&36&	130\\
(11,4)&	10.54&14.9&	1&	0.90&	1.94&1.74	&18&	160\\
(11,6)&	11.70&20.4&	-1&	0.88&	1.56&1.45	&26&	29\\
(11,7)&	12.31&22.7&	1&	0.81&	1.63&1.48	&19&	44\\
(11,9)&	13.59&26.7&	-1&	0.76&	1.45&1.31	&1.8&	--\\
(12,1)&	9.82&	4&	-1&	1.06&	1.64&1.55	&37&	210\\
(12,2)&	10.27&7.6&	1&	0.90&	2.07&1.81	&10&	260\\
(12,4)&	11.30&13.9&	-1&	0.92&	1.55&1.45	&28&	90\\
(12,5)&	11.85&16.6&	1&	0.83&	1.74&1.56	&18&	110\\
(12,7)&	13.04&21.4&	-1&	0.80&	1.47&1.33	&6.2&	--\\
(13,0)&	10.18&0&	1&	0.90&	2.11&1.83	&8.2&	300\\
(13,2)&	11.05&7&	-1&	0.95&	1.54&1.45	&29&	160\\
(13,3)&	11.54&10.2&	1&	0.83&	1.91&1.62	&9.3&	190\\
(13,5)&	12.61&15.6&	-1&	0.83&	1.46&1.34	&8.1&	63\\
(13,6)&	13.18&18&	1&	0.76&	1.61&1.42	&14&	--\\
(14,0)&	10.97&0&	-1&	0.96&	1.53&1.44	&19&	190\\
(14,3)&	12.31&9.5&	-1&	0.86&	1.45&1.35	&1.1&	120\\
(15,1)&	12.16&3.2&	-1&	0.86&	1.45&1.35	&9.3&	160\\
(16,2)&	13.39&5.8&	-1&	0.79&	1.38&1.26	&1.0&	--\\
\end{tabular}
\end{ruledtabular}
\end{table*}
\thispagestyle{empty}


\begin{thebibliography}{17}
\bibitem[{pl()}]{pl}
\bibinfo{note}{In this paper we use photoluminescence to refer to the combined
  information of luminescence and luminescence excitation spectroscopy as is
  normally done in nanotube research. Also PL intensity gerenerally means the
  maximum intensity observed for a particular chirality.}
\end{thebibliography}
\end{document}